\documentclass{mn2e} \usepackage{graphics} \usepackage{epsfig}
\usepackage{color}
\usepackage{xspace}
\usepackage{times}

\def\etal{et\thinspace al.\thinspace}                    %et al.%
\newcommand{\zsun}{Z$_{\odot}$}
\newcommand{\Ha}{\ifmmode {\rm H}\alpha \else H$\alpha$\fi\xspace}
\newcommand{\Hb}{\ifmmode {\rm H}\beta \else H$\beta$\fi\xspace}
\newcommand{\Hg}{\ifmmode {\rm H}\gamma \else H$\gamma$\fi\xspace}
\newcommand{\Hd}{\ifmmode {\rm H}\delta \else H$\delta$\fi\xspace}

\newcommand{\Hii}{\ifmmode \rm{H}\,\textsc{ii} \else H\,{\sc ii}\fi}
\newcommand{\Nii}{[N\,{\sc ii}]$\lambda$6584}
\newcommand{\nii}{\ifmmode [\rm{N}\,\textsc{ii}] \else [N\,{\sc ii}]\fi}

\newcommand{\oi}{\ifmmode [\rm{O}\,\textsc{i}] \else [O\,{\sc i}]\fi}

\newcommand{\neiii}{\ifmmode [\rm{Ne}\,\textsc{iii}] \else [Ne\,{\sc iii}]\fi}

\newcommand{\hei}{\ifmmode [\rm{He}\,\textsc{i}] \else [He\,{\sc i}]\fi}
\newcommand{\oii}{\ifmmode [\rm{O}\,\textsc{ii}] \else [O\,{\sc ii}]\fi}
\newcommand{\Oiii}{[O\,{\sc iii}]$\lambda$5007}

\newcommand{\oiii}{\ifmmode [\rm{O}\,\textsc{iii}] \else [O\,{\sc iii}]\fi}

\newcommand{\sii}{\ifmmode [\rm{S}\,\textsc{ii}] \else [S\,{\sc ii}]\fi}
\newcommand{\siii}{\ifmmode [\rm{S}\,\textsc{iii}] \else [S\,{\sc iii}]\fi}

\title[Can retired galaxies mimic active galaxies?]
      {Can retired galaxies mimic active galaxies?\\ Clues from the Sloan Digital Sky Survey}

\author[Stasi\'nska et al]
         {G. Stasi\'nska$^{1}$, 
	  N. V. Asari$^{2,1}$,
	  R. Cid Fernandes$^{2,1}$,
	  J. M. Gomes$^{2,1}$,
	  	\newauthor
	  M. Schlickmann$^{2}$,
 	  A. Mateus$^{3}$, 
 	   W. Schoenell $^{2}$, 
 	  	  L. Sodr\'e Jr.$^{4}$
 	  	  	  	\newauthor
	(the SEAGal collaboration)\thanks{Semi-Empirical Analysis of Galaxies}\\
	 $^{1}$LUTH, Observatoire de Paris, CNRS, Universit\'e Paris Diderot; Place Jules Janssen 92190 Meudon, France\\
	 $^{2}$Depto.\ de F\'{\i}sica - CFM - Universidade Federal de Santa Catarina,
	 Florian\'opolis, SC, Brazil\\
     $^{3}$Laboratoire dÕAstrophysique de Marseille, CNRS UMR6110, Traverse du Siphon, 13012 Marseille, France \\
	 $^{4}$Instituto de Astronomia, Geof\'{\i}sica e Ci\^encias
         Atmosf\'ericas, Universidade de S\~ao Paulo, S\~ao Paulo, SP,
         Brazil
         }

\begin{document}

\maketitle

\begin{abstract}

The classification of galaxies as star forming or active is generally
done in the (\oiii/\Hb, \nii/\Ha) plane. The Sloan Digital Sky Survey
(SDSS) has revealed that, in this plane, the distribution of galaxies
looks like the two wings of a seagull. Galaxies in the right wing are
referred to as Seyfert/LINERs, leading to the idea that
non-stellar activity in galaxies is a very common phenomenon.  Here,
we argue that a large fraction of the systems in the right wing could actually be galaxies which stopped forming stars. The
ionization in these ``retired'' galaxies would be produced by hot post-AGB stars
and white dwarfs.  Our argumentation
is based on a stellar population analysis of the galaxies via our
STARLIGHT code and on photoionization models using the Lyman continuum
radiation predicted for this population. The proportion of LINER galaxies that can be explained in such a way is however uncertain. We further show how
observational selection effects account for the shape of the right
wing. Our study suggests that nuclear activity may not be  as common as
thought.  If retired galaxies do explain a large part of the seagull's
right wing, some of the work concerning nuclear activity in galaxies,
as inferred from SDSS data, will have to be revised.

\end{abstract}


\section{Introduction}
\label{sec:Introduction}

In the Baldwin, Phillips \& Terlevich (1981, BPT) diagram which is
used to isolate star forming from active galaxies, the
galaxies from the Sloan Digital Sky Survey (SDSS; York et al. 2000)
occupy a well-defined region, evoking the wings of a flying
seagull. The left wing consists of star forming (SF) galaxies while
the right wing is attributed to galaxies with an active nucleus
(Kauffmann et al. 2003, Stasi\'nska et al. 2006). The right wing has
been subdivided into an upper and a lower branch, called ``Seyfert''
and ``LINER'' branch respectively (Kewley et al. 2006). This
denomination is given with reference to the typical emission line
ratios of active galactic \emph {nuclei} (LINER stands for "Low
Ionization Nuclear Emission Regions" in the original paper by Heckman
1980).  However, it is by no means obvious that all galaxies of the
 right wing are powered by an energetically-dominant active
nucleus, i.e., that they are genuine Seyferts or LINERs.  {Binette et al. (1994) found that, in early-type
galaxies that have stopped forming stars, hot post-AGB stars and  white dwarfs provide enough ionizing photons to account for the
observed \Ha\ equivalent widths and can explain the  LINER-like 
 emission line ratios observed in such galaxies 
(see also Sodr\'e \& Stasi\'nska
1999). Taniguchi et al. (2000) followed up this idea, proposing that some LINERs could be post starburst nuclei powered by planetary nebulae central stars.}

 In  Binette et al. (1994), the comparison
with observations was  very limited, due to the scarcity of adequate data at that time. 
With the SDSS, we have a homogeneous data base of over half a million galaxy spectra. In addition, the techniques to extract  emission lines   after modelling the stellar continuum using population synthesis (Kauffmann et al. 2003, Cid Fernandes et al. 2005) now allow emission line measurement even in objects where these  lines are difficult to detect in the raw spectra.

The computations by Binette et al. (1994) and Taniguchi et al. (2000)
were based on \emph {ab initio} stellar population models. Here, we   \emph {directly} use
the  populations inferred from detailed fits of the
observed galaxy spectra to compute
the Lyman continuum radiation and estimate its
impact on the emission lines.
The radiation from the  post-AGB and white dwarf stars present 
in old stellar populations is much harder that the one from young stars, so that galaxies that are not forming stars presently  will contain hotter
emission-line regions, and lie \emph {above} the pure star-forming
sequence in the BPT diagram. In the remaining of this  letter we
qualify these galaxies as \emph {retired} \footnote {\emph {Retired} is to be opposed to  {\em active}, with reference to star formation. We avoid   using the term \emph {passive} since this might suggest ``without emission lines'' (Miller et al., 2003). Ironically, as shown in this letter, a fraction of so-called {\em active  galaxies}, by reference to {\em nuclear activity} (e.g. Kauffmann \& Heckman 2005), could be genuine {\em retired} galaxies.}. 

We explore the expected
properties of retired galaxies  in terms of their line
luminosities and their location in emission line diagnostic diagrams.

\section{Data Base}
\label{sec:Data}

\subsection{Sample and data processing}

This work analyses data extracted from SDSS Data Release 5
(Adelman-McCarthy et al. 2007).  Our parent sample is defined as the
573141 objects spectroscopically classified as galaxies and with no duplicates. We also adopt the following
criteria: $14.5 \le m_r \le 17.77$ (from the definition of the Main
Galaxy Sample), a minimum  $S/N$ of 10 at $\sim 4750$
\AA\ (to ensure a reliable stellar population analysis), $z \ge 0.002$ 
(to avoid intragalactic sources)
and total
z-band light inside the fiber $> 20$ \% (to reduce aperture effects).

The data are processed as in Cid Fernandes et al. (2005) and Mateus et
al. (2006).  The stellar populations composing a galaxy are inferred
through a pixel-by-pixel modelling of its continuum with our code
STARLIGHT, using a base of simple stellar populations computed with
the evolutionary population synthesis code of Bruzual \& Charlot
(2003, BC03).  The emission lines are then measured by fitting gaussians to
the residual spectrum.  We use the same base of 150 stellar
populations with ages 1 Myr $<t_{\star} < 18$ Gyr, and metallicities
$0.005<Z_{\star}/Z\odot<2.5$ as Cid Fernandes et al. (2007).
Illustrative fits are presented in Asari et al. (2007).

\subsection{Chopping the seagull}
\label{sec:chop}

%***FIG***FIG
\begin{figure}
\centerline{
  \includegraphics[scale=0.38]{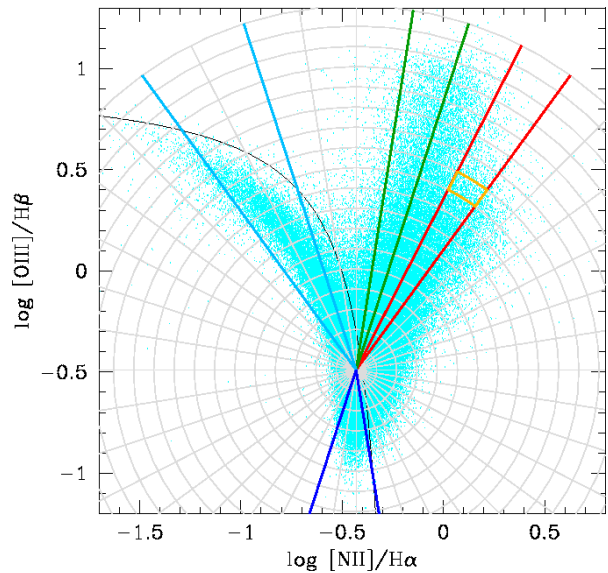}
}
\caption{The chopped seagull and the definition of the SF branches
  (light blue and dark blue), Seyfert branch (green) and LINER branch
  (red). The data points represent all the galaxies defined in
  Sect. 2.1 that have $S/N > 3$ in the 4 lines involved in the plot
  (131287 objects). The black curve is the upper envelope of pure SF
  galaxies according to Stasi\'nska et al. (2006).}
\label{1}
\end{figure}
%***FIG***FIG

Given the distribution of the galaxies in the BPT plane (\oiii/\Hb\ vs
\nii/\Ha), it is convenient to use polar coordinates ($r$, $\theta$),
with the centre defined at the point of inflection of the median curve
for the distribution of \oiii/\Hb\ as a function of \nii/\Ha:
(log \nii/\Ha$=-0.43$; log \oiii/\Hb$=-0.49$). The SF wing corresponds to
$\theta \sim 117\,^{\circ}$ for the low-metallicity branch, and $\sim
-94\,^{\circ}$ for the high-metallicity branch, while the right wing
corresponds to $\theta$ = 45--$90\,^{\circ}$, with $\theta \sim 77
\,^{\circ}$ for Seyferts and $\sim 59 \,^{\circ}$ for LINERs. Emission
line properties, stellar populations and the corresponding ionizing
spectra were obtained after chopping the seagull in 40 angular and 18
radial bins, as shown in Fig.\,1, and analysed for different bins in
$\theta$, as a function of the radial bin index, $i_{\rm r}$.

\section{The stars in the seagull's  wings}
\label{sec:char}

\subsection{Two extreme cases to consider}
\label{sec:prel}

Two versions of the ionizing spectrum are considered for each bin: One
including all the stellar populations inferred from the synthesis
(case F, for ``full'') and another with the contribution of
populations younger than $10^{7.5}$ yr set to zero (case O, for
``old'').  The reason is that, due to errors in the data and
uncertainties inherent to population synthesis, spectral fits of
galaxies without current star formation may attribute a small amount
of light to  young populations.

For instance, along the  LINER branch for $i_{\rm r}>7$, we
obtain light fractions $x_Y$ of 
$\sim 1$\% at 4020 \AA\ for  populations younger than $10^{7.5}$ yr. This corresponds to mass fractions of less than
$10^{-5}$ and is clearly noise.  Yet,
given the 5 orders of magnitude difference between ionizing fluxes of
young and old populations, even such optically insignificant
fractions can have a substantial impact on the shape and intensity of
the radiation field in the Lyman continuum of hydrogen. Setting $x_Y =
0$ circumvents this problem, while at the same time emphasizing the
effects of  post-AGB stars alone.

Along the Seyfert branch, on the other hand, $x_Y$ averages
to 7\%, implying that Seyferts have younger stars than LINERs, in full
agreement with both SDSS (e.g., Kauffmann \etal 2003) and independent
studies (e.g., Gonz\'alez Delgado et al. 2004). All this is clearly
illustrated in Fig.~3 of Cid Fernandes et al. (2008), where we show
how star-formation histories vary across the BPT diagram.

We now ask: what is the impact of the ionizing radiation produced by
the stars responsible for the optical continuum upon  the emission lines in galaxies?

\subsection{The number of ionizing photons}
\label{sec:out}

%***FIG***FIG
\begin{figure}
\centerline{
  \includegraphics[width=0.5\textwidth,bb=18 480 592 718]{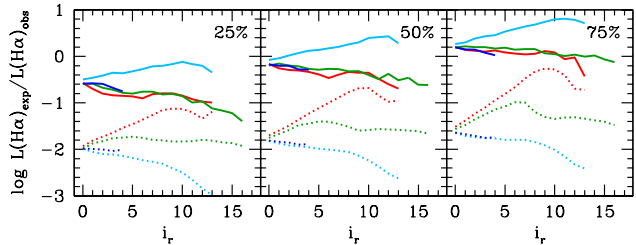}
}
\caption{Variations of $L$(\Ha)$_{\rm exp}/L$(\Ha)$_{\rm obs}$ along
  $i_{\rm r}$ for the upper SF branch (light blue), lower SF branch (dark
  blue), Seyfert branch (green) and LINER branch (red). Full lines:
  case F; dotted lines: case O. The middle panel shows the median
  value of $L$(\Ha)$_{\rm exp}/L$(\Ha)$_{\rm obs}$, while the left and
  right ones show the 25 and 75 quartiles.  }
\label{2}
\end{figure}
%***FIG***FIG

The first question to examine is whether the ionizing photons from the
evolved stellar populations are enough to account for the observed
\Ha\ luminosities in the seagull's wings.  For each galaxy, we compute
 $Q_{\rm H I}$, the number of stellar photons with
energies above 13.6 eV arising from the populations uncovered
with STARLIGHT (we use the BC03 models also in the Lyman
continuum).  We then estimate $L$(\Ha)$_{\rm exp}$, the \Ha\ luminosity expected if all
these ionizing photons are absorbed by the gas present in the
galaxies,  and compare it to  $L$(\Ha)$_{\rm obs}$, the observed value corrected for extinction using H$\alpha$/H$\beta$
and the Cardelli et al. (1989) extinction law for $R_V$=3.1. In each bin of the chopped seagull, we define the median and
quartiles of the $L$(\Ha)$_{\rm exp}/L$(\Ha)$_{\rm obs}$
distributions. In Fig.\,2, we plot their values as a function of
$i_{\rm r}$, with style and colour coding as indicated in the caption.

We see that, in the LINER branch at large radii, the old stellar
populations contribute to  the
ionizing radiation at least as much as the young ones. Given that $x_Y$ 
is very small and thus very uncertain (see Sect. 3.1), old populations could even be dominant. For
the Seyfert branch, on the other hand, it is the young stars which
provide most of the stellar ionizing photons. For the SF branches, the old populations play no role at all. The dispersion
in $L$(\Ha)$_{\rm exp}/L$(\Ha)$_{\rm obs}$ as function of $i_{\rm r}$
can be judged by comparing the curves corresponding to the quartiles
(left and right panels of Fig. 2) with those corresponding to the
median (central panel).

Overall, as seen in Fig. 2, while the stellar populations corresponding to case F can largely explain the upper star-forming branch in terms of total number of ionizing photons, they explain only  $\sim25$\% of the Seyfert and LINER branches. The central panel of Fig. 2 shows  a deficit by a factor 1.5 to 4 for median values of $L$(\Ha)$_{\rm exp}/L$(\Ha)$_{\rm obs}$. In addition, as mentioned in 3.1, the young stellar populations uncovered by STARLIGHT for the LINER branch are not reliable and case O models could be more appropriate, further increasing the discrepancy. However, our plots are plagued with many uncertainties, as will be discussed in Sect.\,5. 

\subsection{The hardness of the ionizing radiation field}
\label{sec:out}

%***FIG***FIG
\begin{figure}
\centerline{
  \includegraphics[width=0.5\textwidth,bb=18 480 592 718]{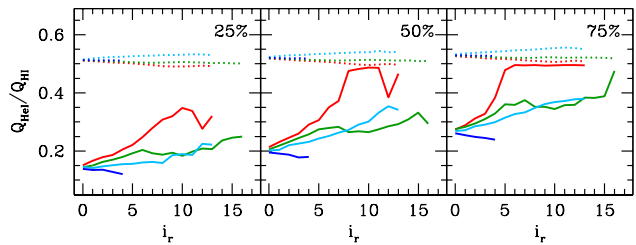}
}
\caption{Variation of $Q_{\rm He I}$/$Q_{\rm H I}$ along $i_{\rm
    r}$. Conventions as in Fig. 2.}
\label{3}
\end{figure}
%***FIG***FIG

%***FIG***FIG
\begin{figure}
\centerline{
  \includegraphics[width=0.5\textwidth,bb=18 480 592 718]{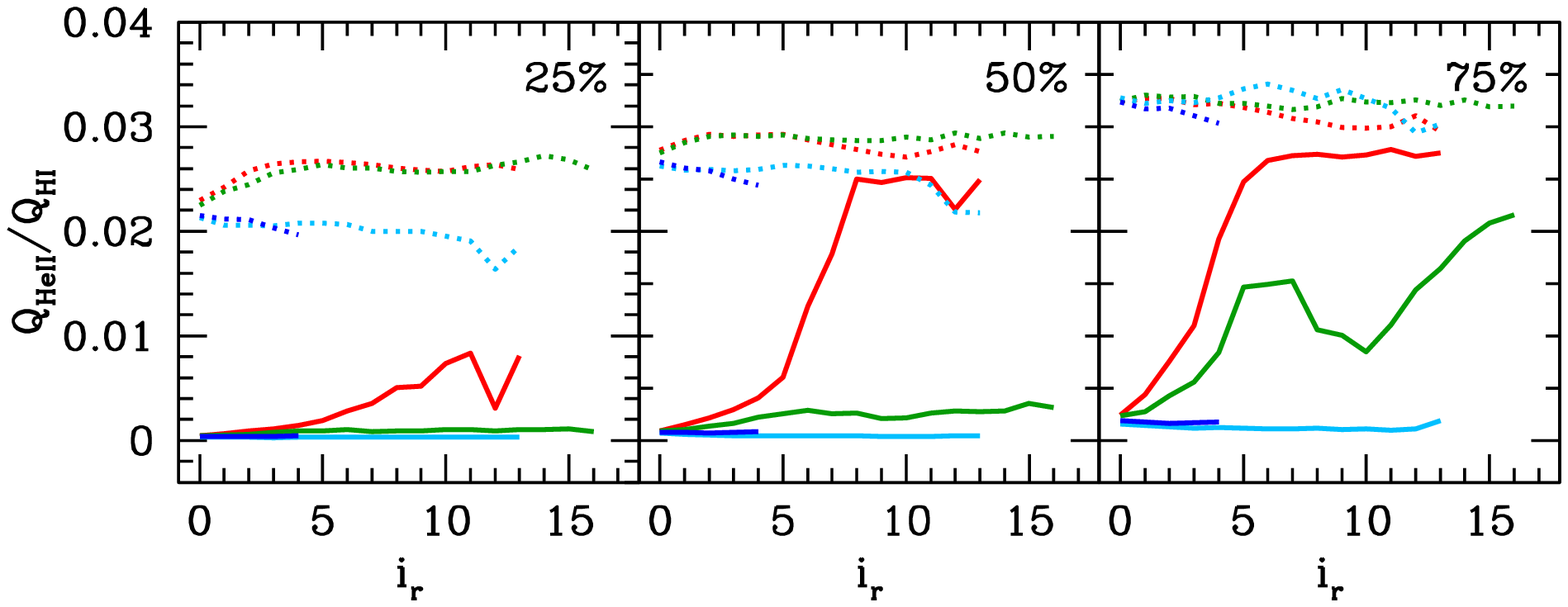}
}
\caption{Variation of $Q_{\rm He II}$/$Q_{\rm H I}$ along $i_{\rm
    r}$. Conventions as in Fig. 2.}
\label{4}
\end{figure}
%***FIG***FIG

The hardness of the ionizing radiation field can be judged by
comparing the values of $Q_{\rm He I}$ and $Q_{\rm He II}$ (the number
of photons above 24.6 eV and 54.4 eV, respectively), to $Q_{\rm H I}$.
Figure 3 shows $ Q_{\rm He I} / Q_{\rm H I}$ as a function of $i_{\rm
  r}$ for the same values of $\theta$ as in Fig.\ 2.  Again, we show
curves for galaxies whose $Q_{\rm He I} / Q_{\rm H I}$ values
correspond to the quartiles (left and right panels) and the median
(central panel) of the distribution of $Q_{\rm He I} / Q_{\rm H I}$
for each branch. It is clearly seen that the radiation produced by the
stellar populations in the LINER branch is much harder than in the SF
and Seyfert branches, especially for $i_{\rm r}$ $>5$. Note also that
the curves for Seyfert and upper SF branches are similar. This is
likely the result of a ``cosmic conspiracy''. While, for the SF
branch, the hardening of the radiation is due to the decrease of
metallicity (McGaugh 1991), the origin of the hardening in the Seyfert
branch stems from an increasing population of old stars (see Fig. 3 of Cid Fernandes et al. 2008).  As expected, the curves for case O
indicate harder radiation, and reach a maximum equal to the typical
value corresponding to evolved stellar populations containing hot
post-AGB stars and white dwarfs.

Fig. 4 is is analogous to Fig. 3 for $Q_{\rm He II} / Q_{\rm H I}$ and
shows a very different behaviour of the LINER branch with respect to
the other branches, since the effect of hot post-AGB
stars and white dwarfs  is here dominant.  Note that the values of $Q_{\rm He II}$ do
not take into account the X-ray radiation produced by hot stars as well
as by X-ray binaries, both of which could have a non-negligible
contribution to $Q_{\rm He II} / Q_{\rm H I}$.

\section{Photoionization models for the right wing}
\label{sec:phot}

In photoionized nebulae, the emission line ratios are basically
determined by three parameters: the hardness of the ionizing
radiation, the nebular metallicity and the ionization parameter $U$
(defined as $Q_{\rm H I}/(4 \pi R^2 n c)$ where $R$ is nebular radius,
$n$ is the gas density and $c$ is the speed of light).  We present
photoionization models using the ionizing radiation from the stellar
populations corresponding to Figs. 3 and 4 for case F. Not much is known 
about the gas distribution in the galaxies of the right wing. For
simplicity, we have assumed a thin shell geometry, with density
$n=500$ cm$^{-3}$.  Using the code PHOTO, we have computed models with
different values of $U$ and of the nebular metallicity $Z$ (defined as
the oxygen abundance in units of $4.9 \times 10^{-4}$, the solar value from
Allende Prieto et al. 2001). The abundances of the other elements
follow the same prescriptions as in Stasi\'nska et al. (2006).

We first discuss the LINER branch.  In Fig. 5, we show the location of
several model sequences in classical line ratio diagrams,
superimposed on the data points. Each sequence has $Z$ varying from
0.03 to 6 \zsun, and is defined by $U$ as indicated in the figure
caption. The ionizing radiation for all the models is given by the
stellar population of the galaxy having the median value of $Q_{\rm He
  I} / Q_{\rm H I}$ in the bin marked in orange in Fig. 1. Note that
the results are almost the same when changing the radiation field from
$i_{\rm r}$=7 upwards in the LINER branch.  We see in Fig. 5 that
models with metallicities twice solar cover the tip of the LINER
branch in the BPT diagram, provided that log $U$ is between -3 and
-4. Models with even higher $Z$ can be found in this same region. With
the softer radiation field corresponding to smaller values of $i_{\rm
  r}$, one produces models that can cover the inner part of the LINER
branch.  We conclude that the radiation from old stellar population in
metal rich galaxies can easily account for  the emission line ratios observed in the LINER
branch. The other diagrams shown in Fig. 5 are also in reasonable
agreement with this proposition.

We now turn to the Seyfert branch. Here, the ionizing radiation field
provided by the evolved stellar populations is significantly softer
than in the LINER branch, as seen from Figs. 3 and 4. Therefore, one
can infer from Fig. 5 that the emission line ratios for the upper Seyfert branch cannot be
explained by stellar radiation alone.

%***FIG***FIG
\begin{figure}
\centerline{
  \includegraphics[scale=0.38]{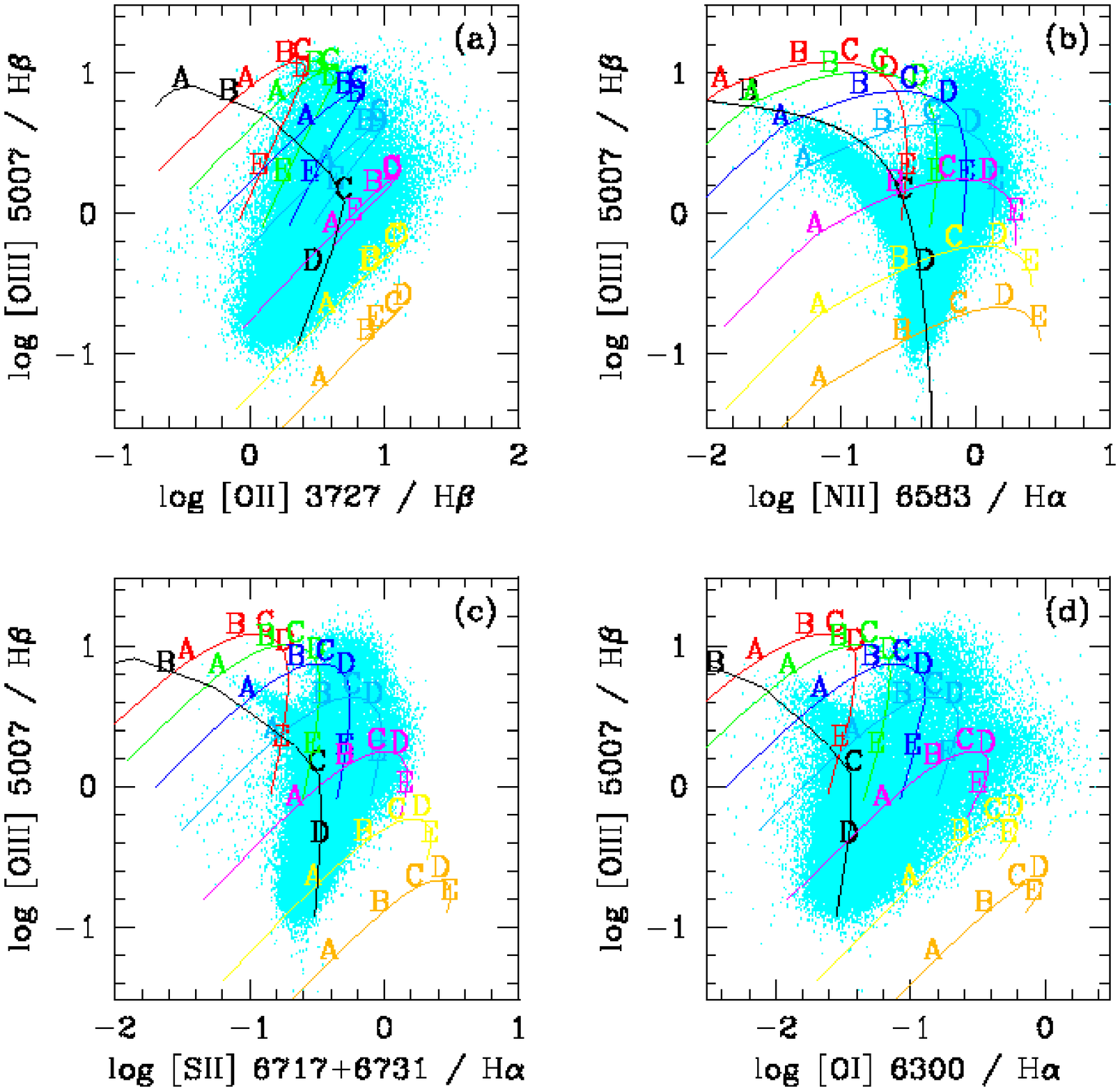}
}
\caption{Our sample galaxies in four classical emission line ratio
  diagrams: \oiii/\Hb\ versus: \oii/\Hb\ (a), \nii/\Ha\ (b),
  \sii/\Ha\ (c) and \oi/\Hb\ (d).  Model sequences for the LINER
  branch are shown for different values of log $U$: -2.3 (red), -2.7
  (green), -3 (blue), -3.3 (cyan), -3.7 (purple), -4 (yellow), -4.4
  (orange). The black line is the model sequence for SF galaxies from
  Stasi\'nska et al. (2006). The metallicities $Z/Z_\odot$ are marked
  with letters as follows: 0.2 (A), 0.5 (B), 1 (C), 2 (D), 5 (E).}
\label{5}
\end{figure}

\section{Uncertainties}
\label{sec:un}

As has just been shown, our  models indicate that the  old stellar populations responsible for the optical continua of the considered galaxies can explain the observed emission line ratios of  the LINER branch. However  the observed \Ha\ luminosities  are more difficult to explain. Our case O models reproduce $L$(\Ha)$_{\rm obs}$ within a factor 2 for about 25\% of the LINER galaxies with $i_{\rm
  r}$ $> 7$\footnote{For $i_{\rm
  r}$ $< 7$, the contribution of young stellar populations found by STARLIGHT is likely real, and $L$(\Ha)$_{\rm obs}$ is easily reproduced by case F.}.  Does this mean that about one quarter of the LINER galaxies are ionized by hot post-AGB and white dwarf stars, the rest being powered by some other source? In view of the many uncertainties involved in this study, such a conclusion would be premature.    

First,  there are already uncertainties in the mere process of population synthesis fitting of galaxy spectra (see Cid Fernandes et al. 2005). Second, the applied extinction correction may not be appropriate. 

Perhaps the most important point is that the  modelling of the Lyman continuum in old stellar populations is very uncertain. Unfortunately, ready-to-use evolutionary population synthesis codes that compute the ionizing radiation from hot post-AGB and white dwarf stars are scarce. We made experiments with the BC03 code and with PEGASE (Fioc et al. 1997). With BC03, for an instantaneous starburst, we find that  $Q_{\rm H I}$ 
depends on the metallicity: it increases by about 0.3\,dex from $Z$ = \zsun\ to 0.02\,\zsun\ at ages larger than $10^8$\,yr, and by over one order of magnitude at around  $10^8$\,yr. Therefore, any mistakenly assigned   metallicity  during the continuum fitting process will induce some error on $Q_{\rm H I}$. With PEGASE, the values of $Q_{\rm H I}$ for solar metallicity are larger by 0.2--1 dex than with BC03. What is the reason for such a difference? Different physical ingredients (stellar evolutionary tracks and atmospheres)? Or inaccuracies in the numerical treatments? It  seems to us that the major effect comes from the initial-final mass relation of white dwarfs. The analytic form of this relation has been  revised recently (Catal\'an et al 2008), which will certainly lead to changes in the $Q_{\rm H I}$ predictions. However, the observational dispersion is very large (see their Fig. 2), so that the predictions will remain uncertain until one understands better the drivers of the initial-final mass relation(s).

\section{Why we see a seagull in the BPT diagram}
\label{sec:seagull}

By merging the results of Sects. 3 \& 4 with those of Stasi\'nska et al. (2006),
we find that photoionization models built with realistic stellar
populations are able to cover nearly the entire BPT plane. We now
discuss why some regions are avoided by real galaxies.

SF galaxies form a narrow wing because, as shown by Dopita et
al. (2006), in regions of star formation the metallicity controls both
the ionizing radiation field and the ionization parameter.

But why is the right wing also narrow? Our models for retired galaxies
cover the entire space between the two wings of the seagull as well as
below the wings.  One factor which plays a role in shaping the seagull
is that, in order to appear in the BPT diagram, galaxies must be
sufficiently bright in the $r$ band to satisfy the SDSS selection 
criteria for spectroscopy, and must have the 4 involved emission lines measured with
sufficient $S/N$.  As a matter of fact, the lower border of the right
wing is traced by the galaxies with the lowest \oiii\ equivalent
widths in our sample.

%***FIG***FIG
\begin{figure}
\centerline{
  \includegraphics[width=0.52\textwidth, bb=18 520 592 700]{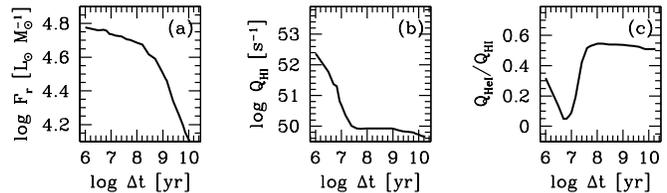}
}
\caption{Evolution of (a) the $r$-band flux, (b) $Q_{\rm H I}$ and (c)
  $Q_{\rm He I}$/$Q_{\rm H I}$ for the galaxy corresponding to the median
  of stellar age in bin $i_r = 8$ of the upper SF branch.}
\label{6}
\end{figure}
%***FIG***FIG

Metal poor retired galaxies (models A and B in Fig. \,5) should be
found below the SF wing and between the wings, if they exist and can
be detected.   
As shown in Cid Fernandes et al. (2007), low metallicity SF galaxies are forming stars efficiently (the ``downsizing'' phenomenon), so that our retired low $Z$ galaxies (models A and B in Fig. 5) may just be ``ahead of their time''. Even if such galaxies existed, they would be faint both in
their emission lines and stellar continuum.  To illustrate this, we
predict the future evolution of typical low metallicity SF galaxies under the
assumption that they stop forming stars now. Applying the models of
BC03 to their present-day stellar populations, we
find that 0.1 Gyr from now their radiative output will have dropped by
a factor of $\sim 300$--$1000$ at $h\nu > 13.6$ eV, and in 1 Gyr their
$r$-band flux will have faded by $\sim 0.4$--$0.8$ mag (Fig.~6).  From the observed
$m_r$ distribution of the metal poor SF galaxies in the SDSS, we
estimate that over 90\% of them should fade beyond the $m_r < 17.77$
limit by the time they retire.  Hence, metal poor retired galaxies
either do not exist yet or are too faint, which eliminates all the
models below the left wing and between the two wings in Fig 5.

These considerations imply that detectable retired galaxies must have
metallicities of the order of solar or larger, and be massive.   As
shown in Fig. 7a and b, this is in agreement with the mean stellar
metallicities ($Z_\star$) and masses ($M_\star$) obtained  using STARLIGHT for
those galaxies.

Finally, why are there no SF galaxies with such high metallicities in
the BPT diagram? The first idea that comes to mind is that, because of
downsizing, such galaxies have already stopped forming stars.  There
is another possibility, of different nature: nebular metallicities
above $\sim 3$ Z$_\odot$ lower the electron temperature so much
that  \Oiii\ cannot be excited (this does
not happen in the right wing, since the harder ionizing radiation
field produces more efficient heating and compensates the cooling
caused by the large metal abundance).  Interestingly, there are
emission-line galaxies in the SDSS that could  well be SF  with metallicities larger than 2--3 Z$_\odot$, i.e. the ancestors of the most metal-rich retired galaxies: they have
log \nii/\Ha$<-0.3$ and no \oiii.  In Fig. 7c, we plot these galaxies
(in black) in a $M_\star$ vs $Z_\star$ diagram, together with the
other SF galaxies (in blue).  The black points populate the zone of
high $M_{\star}$ and $Z_\star$, in agreement with our interpretation.
It is their high metal content which excludes these likely
ancestors of metal rich retired galaxies from the BPT diagram.

%***FIG***FIG
\begin{figure}
\centerline{
  \includegraphics[width=0.42\textwidth]{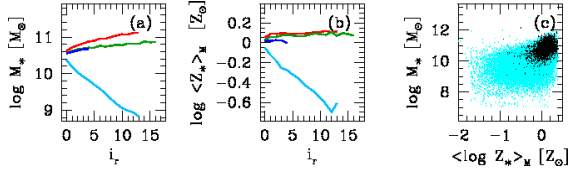}
}
\caption{The stellar mass $M_{\star}$ (a) and mean stellar metallicity
  $Z_{\star}$ (b) along $i_{\rm r}$, same conventions as Fig. 2. (c)
  $M_{\star}$ vs. $Z_{\star}$; cyan: SF galaxies from the
  BPT diagram, black: galaxies with no \oiii\ and log \nii/\Ha$<-0.3$.}
\label{7}
\end{figure}
%***FIG***FIG

Note that the concept of retired galaxies applies also to galaxies that belong to the red sequence studied by Graves et al. (2007) and have such a weak \Hb\ that they do not appear in the BPT diagram, and  \Nii/Ha ratios similar to those of LINER galaxies. As a matter of fact, there is a numerous population of such galaxies in the SDSS (over 40,000 using the selection criteria of Sect. 2.1). The weaker \Ha\ equivalent widths of these galaxies  are easily reproduced by our models for retired galaxies.

\section{Conclusion}

We have shown that retired galaxies can account for a good part of the 
LINER branch in the seagull's right wing. How much exactly is difficult to say, given the uncertainties.  
If, as we argue, a significant fraction of the LINER galaxies are in fact retired galaxies -- and not active galaxies as
generally claimed -- the perception of the local Universe would be
drastically changed. Nuclear activity would not be as common as
thought, and  some of the work based on SDSS data and related to the AGN
population will have to be reconsidered.

\section*{Acknowledgements}

This work was supported by the CAPES-COFECUB program.  The
STARLIGHT project is supported by the Brazilian agencies CNPq, CAPES
and FAPESP.  The Sloan Digital Sky Survey is a joint project of The
University of Chicago, Fermilab, the Institute for Advanced Study, the
Japan Participation Group, the Johns Hopkins University, the Los
Alamos National Laboratory, the Max-Planck-Institute for Astronomy, the Max-Planck-Institute for Astrophysics, New Mexico
State University, Princeton University, the United States Naval
Observatory, and the University of Washington.  Funding for the
project has been provided by the Alfred P. Sloan Foundation, the
Participating Institutions, the National Aeronautics and Space
Administration, the National Science Foundation, the U.S. Department
of Energy, the Japanese Monbukagakusho, and the Max Planck Society. We thank the referee for insightful comments.

\end{document}